# AN EFFICIENT APPROACH FOR GENERALIZED LOAD BALANCING IN MULTIPATH PACKET SWITCHED NETWORKS

G. G. Md. Nawaz Ali[1,2], Rajib Chakraborty[2], Md. Shihabul Alam[2] and Edward Chan[1]

[1]City University of Hong Kong, Hong Kong, China
taposh_kuet20@yahoo.com, csedchan@cityu.edu.hk
[2]Khulna University of Engineering & Technology, Khulna, Bangladesh
rajib_csedept@yahoo.co.uk, alam_shihabul@yahoo.com

## ABSTRACT

*This paper is a quantitative analysis on packet switched network with a view to generalize load balancing and determination of appropriate routing algorithm in multipath environment. Several routing algorithms have been introduced for routing of packets from source to destination. Some of them route packets accurately with increased workload and some of them drastically cut down the workload. A few of them can find out a minimum workload deviation for both UDP and TCP packets. We simulated these approaches in a well defined simulator, analyzed and evaluated their performance. After expanding our analysis with varying weights and number of paths we found that the recently proposed routing algorithm Mixed Weighted Fair Routing (MWFR) outperforms the existing routing algorithms by reducing the routing and network overhead and saving the scarce bandwidth as well as CPU consumption for packet switching networks.*

## 1. INTRODUCTION

Load sharing is an important factor in multipath communication networks to optimize bandwidth in this modern era of communication. Now-a-days, the convergence of the computer, communications, entertainments and consumer electronics industry is driving an explosive growth in multimedia applications [4]. Now ISPs are confronted to provide the quality of service (QoS) due to the huge development of internet based multimedia applications. To meet this capacity expansion one efficient solution is to install new links in parallel with existing ones. This requires an effective approach for routing and distributing huge volume of traffic through a set of parallel links. There are some unipath routing protocols [4][5] which can adapt very quickly to changing network conditions but they become unstable under heavy load and bursty traffic conditions and at any given time some subnets can be heavily congested whereas others remain under-utilized.

In this paper, performances of different routing algorithms are measured to determine their effectiveness in performing load sharing in multipath communication networks. Here a generalized load sharing (GLS) [1] [3] mode has been implemented to conceptualize how traffic is split ideally on a set of ideal paths. A simple traffic splitting algorithm, called Weighted Fair Routing (WFR) [1], has been implemented at two different granularity levels, namely the packet level and the call level, to approximate GLS with the given routing weight vector.

The packet by packet WFR (PWFR) [1] mimics GLS by transmitting each packet as a whole, whereas the call by call WFR (CWFR) [1] [2] imitates GLS so that all packets belonging to a single flow are sent on the same path. Here we also implemented a hybrid algorithm which can handle both packet level and call level flow of traffic. We have developed some performance bounds for the PWFR, CWFR and hybrid algorithm and found that PWFR and hybrid algorithms are deterministically fair traffic splitting algorithm. The ultimate goals of this paper are-

- To invoke a new routing algorithm for the packet switching networks that will route efficiently all the packets.
- To drastically minimize workload deviation in the network and hence enhance the performance of the packet switching networks.
- To reduce the routing and network overhead in order to save scarce bandwidth and CPU consumption for packet switching networks.
- To simulate the proposed approach using NS-2 and evaluate its performance against existing efficient routing protocols and proves the proposed routing protocol outperforms existing routing protocols.

## 2. ROUTING APPROACHES

### 2.1 Round Robin (RR) Approach

The most common form of traffic splitting is to distribute packets on a set of active paths in a round robin fashion (RR) [1] or dispense bursts of packet to all participating paths in a cyclic manner. This algorithm is based on the number of packets and not the size of the packets. Since packets generally have different sizes in packet switched networks (except ATM networks), so actual workloads deviate unboundedly from the expected workloads. Although such an algorithm is not difficult to implement, it can only support uniform traffic splitting or cyclic dispersion on these active paths.

We propose to use the RR routing approach for multipath communication using the routing weight vectors assigned to the paths of a node's outgoing links to generalize the load sharing among the paths. The Generalized RR (GRR) approach distributes packets through its outgoing links in a manner so that number of packets or calls in a path relative to the sum of all paths is as close as possible to its routing weight vector.

#### 2.1.1 Packet Based Generalized RR (PGRR)

Packet based RR works on packets. It takes a whole packet as a unit irrespective of size of the packet. PGRR splits packets to each path in a cyclic manner and tries to uniform the arrival instants of any two packets of a path. This approach is suitable for connectionless traffic (UDP traffic). The PGRR algorithm is given in Figure 1.

#### 2.1.2 Call Based Generalized RR (CGRR)

Call-connection based RR deals with number of connection rather than number of packets because this technique is suitable for connection-oriented traffic (TCP flow). It dispatches a set of incoming calls to a set of outgoing calls so that the resulting load distribution in terms of number of calls would be close to a path's weight vector. The complete algorithm is in Figure 2.

```
Monitor the QUEUE(Qi) for all outgoing links of the splitter node.
Define the PROCEDURE(PROC_P) which hold the
link identification number at the time of data transfer.

PROC PROC_P(Path)
BEGIN
        if( path >=N)
                set 1 to path (start from initial)
        else
                set path = path+1
        RETURN path
END
PROC PRR_Packet (Packet)
BEGIN
        set path i to 0
        determine the path i by calling PROC_P(i) for every link
        i ← Packet.size
        Place packet to the output queue of path i
END.
While running time exist call PRR_Packet(Packet)
```

Figure 1. Packet based generalized RR algorithm

```
Monitor the QUEUE(Qi) for all outgoing links of the splitter node.
Define the PROCEDURE(PROC_P) which hold the
link identification number at the time of data transfer.

PROC PROC_P(Path)
BEGIN
        if( path >=N)
        set 1 to path (start from initial)
        else
        set path = path+1
        RETURN path
END
PROC CRR_Packet (Packet)
BEGIN
        set path i to 0
        determine the path i by calling PROC_P(i) for every link
        i ← Packet.Call
        For the same connection place all packets to the output queue
        of path i
END.
While running time exist call CRR_Packet(Packet)
```

Figure 2. Call based generalized RR algorithm

## 2.2 Weighted Fair Routing Approach (WFR)

The two most common internet transport protocols are TCP and UDP. Each TCP connection requires its packets to arrive at the destination in order. If a TCP connection, routes packets on multipath simultaneously, those packets sent on different paths may arrive at the destination out of order. Packet based load sharing approaches may not work well for TCP flows and other connection oriented flows that requires packets at the destination in order. Yet, a call-based multiple path routing approach can be applied for load sharing. A UDP connection or any other connectionless traffic allows packets to arrive at the destination out of order, without affecting protocol performance.

Because of the above requirement, a load sharing approach, called Weighted Fair Routing (WFR) has been proposed [1][2][3]. The packet by packet WFR (PWFR) is a packet level WFR in which a set of packets is split on a set of outgoing channels or links and sent the packet as a whole whereas the call by call WFR (CWFR) is a call-connection level WFR in which a set of connections is split on a set of outgoing channels and all packets belonging to the same connection are routed on the same path.

### 2.2.1 Packet Based WFR (PWFR)

Suppose there is a sequence of packet, namely, packet1, packet2 …, to be split on a set of N paths or channels. Denote the size of packet k by S(k) bytes. The routing weight for path i is given as $p_i$, where

$$\sum_{i=1}^{N} Pi = 1.$$

Define the routing weight vector as P = ($p_1$ $p_2$ … $p_N$) and assume $W_i^p$ and $\hat{W}_i^p$ be the expected and actual workload in byes to be sent on path i.

A metric is introduced to measure the traffic under load on a path. The residual workload of path i, where i = 1, 2… N, just before the routing decision for packet k is made, $R_i^P(K)$ is defined as the amount of work that should be fed on path i in order to achieve the expected workload $W_i^P(K)$. According to [1][3]

$$R_i^p(K) = \{ W_i^p(1) \quad \text{if } k = 1;$$
$$\{ W_i^p(K) - \hat{W}_i^p(K-1) \quad \text{otherwise}$$

$R_i^P(K)$ is used to measure the traffic under load on path i, just before the routing decision of packet k is made. The $K^{th}$ packet will be set to path i depends on the value of $R_i^P(k)$.

If $R_i^P(K) > 0$, path i has been injected with less traffic than expected, and packet k can be sent on this path. On the other hand, if $R_i^P(K) < 0$, path i has too much traffic being routed on it and hence packet k should not be transmitted on this path. If there still path selection problem exist, largest routing weight vector and after that smallest path identification number would be the key for selection. Figure 3 exhibits the complete PWFR algorithm.

```
Monitor the QUEUE(Qi) for all outgoing links of the splitter node.
Define the PROCEDURE(PROC_i) which hold the link
bandwidth at the time of data transfer.

PROC PROC_i(Path)
BEGIN
        S ← Packet. Size
        For each path i from 1 to N
        R_i^P ← R_i^P + P_i S
        Choose a path j such that R_j^P is maximized
        RETURN R_j^P
END

Determine the bandwidth after sending data

PROC TERMINATION (packet)
BEGIN
        R_j^P ← R_j^P – S
END.

PROC PWFR_PACKET (Packet)
BEGIN
        determine the path i which has largest bandwidth by calling
        PROC_i from every link
        i ← Packet. Call. packet
        Place packet to the output queue of path i
        call PROC TERMINATION  after sending data.
END.
While running time exist call PWFR_PACKET(Packet)
```

Figure 3: Packet based WFR algorithm

### 2.2.2 Call-connection Based WFR (WFR)

Using PWFR for the same sequence of packets in the multipath environment creates a performance decline as packets may arrive to the destination out of order. For handling these connection oriented packets a flow-based algorithm instead of packet based algorithm is needed.

Assume-, each call has a finite and average bandwidth requirements and connection k needs a bandwidth requirement of Q(k) units from the sending node whereas the node has N outgoing channels for making an outgoing call.

$\hat{w}_i^c(k)$ and $w_i^c(k)$ are the reserved and expected bandwidth on path i just before connection k is established and when the connection k is made respectively. The total reserved bandwidth for all calls including the incoming call k, on all outgoing channels is, according to [2][3]

$$A(k) = \sum_{i=1}^{N} \hat{w}_i^c(k) + Q(k) \text{ and } w_i^c = P_i.A(k)$$

where, P = (p_1 p_2 … p_N) and i = 1,2,……..N

Thus, the bandwidth deviation $R_i^C(k)$ which is defined as the amount of bandwidth that should be reserved on path i in order to have a reserved bandwidth equal to the expected reserved bandwidth on path i is

$$R_i^C(k) = w_i^c(k) - \hat{w}_i^c(k)$$

If $R_i^C(k)>0$, path i is underloaded and connection k can be routed on this path; on the other hand if $R_i^C(k)<0$, path i is already overloaded and connection k should not be carried on path i. The path which has the maximum $R_i^C(k)$ value is chosen for sending the $k^{th}$ packet. The same procedure is applied as PWFR if there is a tie. The complete CWFR algorithm is summarized in Figure 4.

---

**Monitor the QUEUE(Qi) for all outgoing links of the splitter node.
Define the PROCEDURE(PROC_i) which hold the link
bandwidth at the time of data transfer.**

**PROC PROC_i(Path)**
**BEGIN**
    Qi ← Call.Bandwidth
    A ← $\sum_{i=1}^{N} \hat{w}_i^c$ + Q
    For each path i from 1 to N
    $w_i^c$ ← $P_i$ A
    $R_i^C$ ← $w_i^c - \hat{w}_i^c$
    **RETURN $R_i^C$**
**END**

**Determine the bandwidth after sending data**

**PROC TERMINATION (call)**
**BEGIN**
    Q ← Call. Bandwidth
    i ← Call. Path
    $\hat{w}_i^c$ ← $\hat{w}_i^c$ - Q
**END.**

**PROC CWFR_PACKET (Packet)**
**BEGIN**
    determine the path i which has largest bandwidth by calling
    **PROC_i** from every link
    i ← Packet. Call. Path
    Place packet to the output queue of path i
    $\hat{w}_i^c$ ← $\hat{w}_i^c$ + Q
    call **PROC TERMINATION** after sending data.
**END.**
**While running time exist call CWFR_PACKET(Packet)**

Figure 4. Call based WFR algorithm

## 3. SIMULATION, ANALYSIS AND PERFORMANCE EVALUATION

### 3.1 Performance Metric

We measure the performances of different routing algorithms with respect to workload deviation against routing weight vectors of the paths and time. Workload deviation is a measurement which estimates how the actual workload varies from the expected workload of a traffic splitter of its N possible outgoing paths.

Assume-, M is a set of packets of the incoming traffic of a traffic splitter, where the traffic may be connection-oriented, connectionless or mixed.

Let $w_i^p(k)$ and $\hat{w}_i^p(k)$ be the expected and actual workloads in bytes assigned respectively to path i after the routing decision for the $K^{th}$ packet has been made. The mean square workload deviation of a traffic splitter for M set of packets and N outgoing link is [1][2][3]

$$E\left[(\hat{w}^p - w^p)^2\right] = \frac{\sum_{k=1}^{M} \sum_{i=1}^{N} \left(\hat{w}_i^p(k) - w_i^p(k)\right)^2}{MN}$$

### 3.2 Simulation and Analysis of PWFR, CWFR and GRR

We have implemented the simulation experiments in our research work using Networking Simulator-2 (NS-2) [6][7]. For our simulation we have generated three and five multipath network environment by changing network topology and used those networks extensively to analyse the GRR, PWFR, CWFR and MWFR routing algorithms to determine which algorithm has the best performance for a variety of traffic types (i. e. TCP, UDP and Mixed).

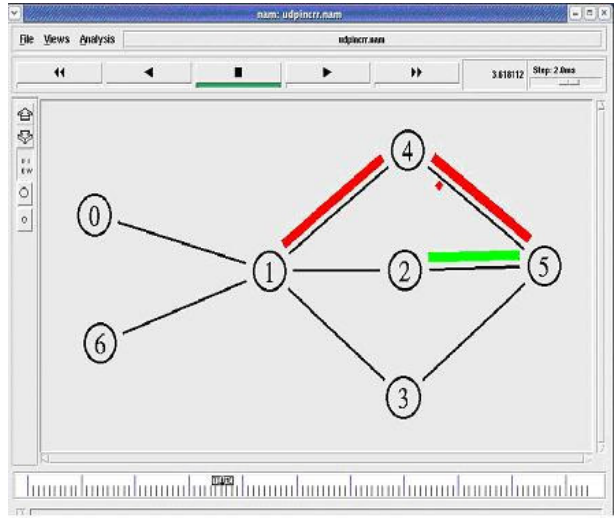

Figure 5. Simulated multipath network when number of path is 3

The routing weighted vector p = (p1, p2, p3), has been set such that p3 = 0. When p3 = 0, p1 varies between 0.001 and 0.5. Due to symmetry, it is not necessary to perform duplicate experiment when p1 is greater than 0.5. The result compares the effectiveness of different traffic splitter under three different traffics types, namely, connection oriented, connectionless and mixed traffic. Figure 6 shows the mean square workload deviation for connectionless traffic when the routing weight for path 3 is 0. Here it is seen that the mean square workload deviation when WFR is employed is significantly lower than GRR.

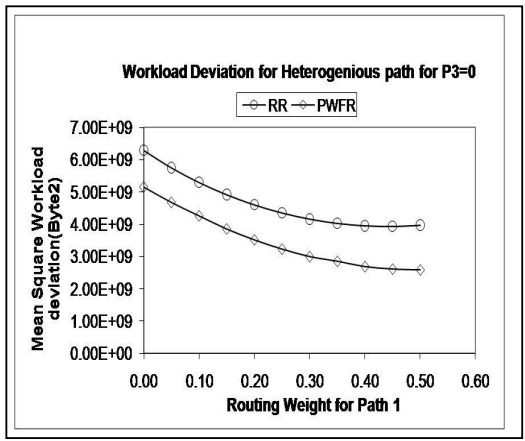 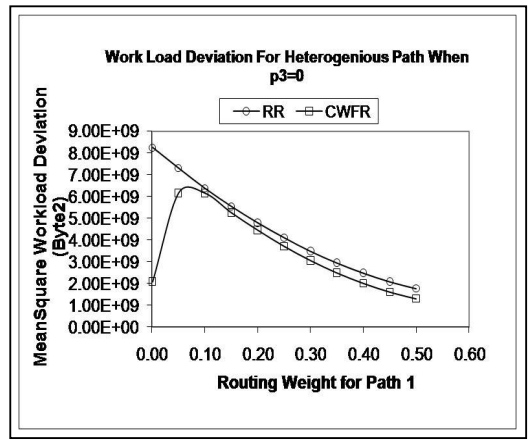

Figure 6. Comparison between GRR and PWFR for connectionless traffic

Figure 7. Comparison between GRR and CWFR for connection-oriented traffic

Again Figure 7 shows the mean square workload deviation for connection oriented traffic when the routing weight for path 3 is 0. Here we also found-, the mean square workload deviation when WFR is employed is significantly lower than GRR. Compared with the cases for connectionless traffic, its superiority fades as all traffic within a single call must be transmitted on the same path once determined.

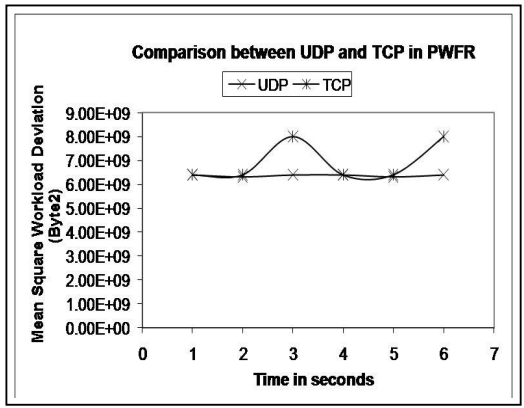 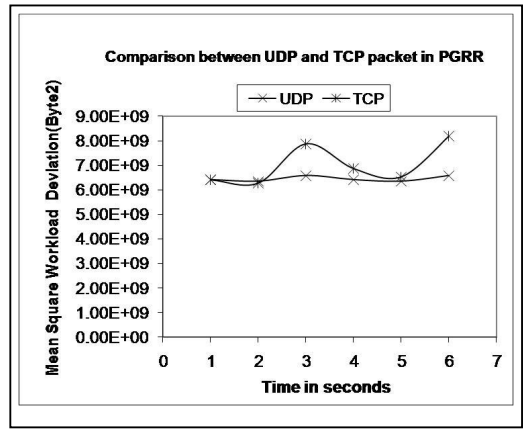

Figure 8. Comparison graph when applying both UDP and TCP in PWFR

Figure 9. Comparison graph when applying both UDP and TCP in PGRR

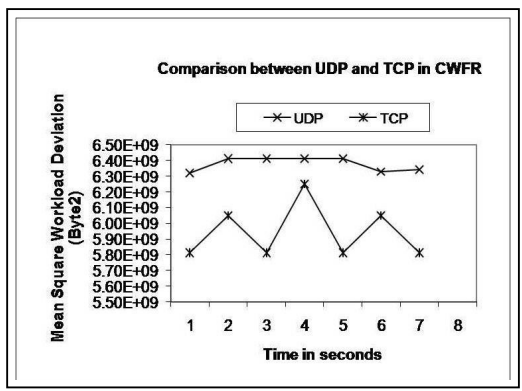 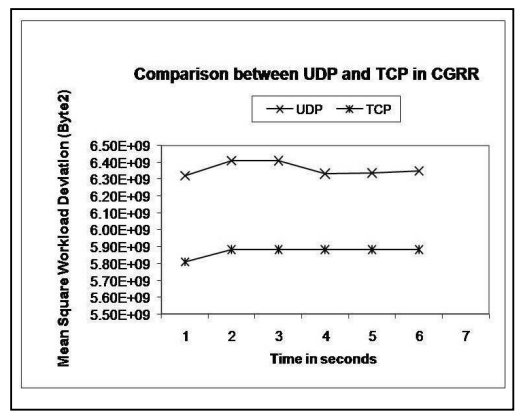

Figure 10. Comparison graph when applying both UDP and TCP in CWFR

Figure 11. Comparison graph when applying both UDP and TCP in CGRR

149

### 3.3 Necessity of Combined Algorithm

Generally, packet-switching networks need to handle both connection-oriented and connectionless traffic. We may need to couple the PWFR and CWFR algorithms to obtain the Mixed WFR algorithm. When a packet is to be forwarded to another node, it is necessary to determine whether it belongs to connection less traffic or connection-oriented traffic. If the packet is connectionless then PWFR is required to handle it, else CWFR is called because of its superiority in splitting UDP and TCP packets by PWFR and CWFR respectively.

Figure 8 compares results for PWFR when UDP and TCP packets are applied and it demonstrates again that UDP is appropriate for PWFR. Figure 9 shows the results for PGRR when UDP and TCP packets are applied and also shows clearly that UDP is appropriate for PGRR.

On the other hand, Figure 10 shows that TCP is appropriate for CWFR. In Figure 11 we found that after applying both types of packets (TCP and UDP) in CGRR, TCP is appropriate for CGRR.

So conclude that the node which would act as a splitter should use both the PWFR and CWFR algorithm and a suitable mechanism for invoking the appropriate algorithm after the type of the packets has been determined. Figure 12 is a summary of Mixed WFR (MWFR) algorithm which handles both connectionless and connection-oriented traffic.

```
Monitor the QUEUE(Qi) for all outgoing links of the splitter node.
Define the PROCEDURE(PROC_i) which hold the link
bandwidth at the time of data transfer.

        PROC MWFR_Packet (packet)
        BEGIN
                If (packet. class is connectionless)
                        call PWFR_Packet (Packet)
                Else
                        call CWFR_Packet (Packet)
        END.
While running time exist  call MWFR_Packet(Packet)
```

Figure 12. Mixed WFR algorithm

### 3.4 Performance Comparison among PWFR, CWFR, GRR and MWFR

Mixed traffic is composed of both connection oriented and connectionless traffic. Figure 13 is the comparison between PWFR, CWFR and MWFR algorithm for mixed traffic. Here, it is clearly seen that MWFR is better than all other WFR traffic splitting algorithms. Again, Figure 14 shows the performance comparison among PGRR, CGRR and GRR. Here, we also find that mixed traffic has a better performance when handled by Mixed RR than PGRR or CGRR.

Figure 15 shows the mean square workload deviation for connectionless traffic when the routing weight for path 3 is 0 i.e. no traffic will be routed on path 3. The routing weight for path 1 varies from 0.001 to 0.5, so opposite consequence for path 2. Here it is seen that the mean square workload deviation of MWFR is always lower than that of GRR that is MWFR outperforms GRR or Mixed RR (MRR).

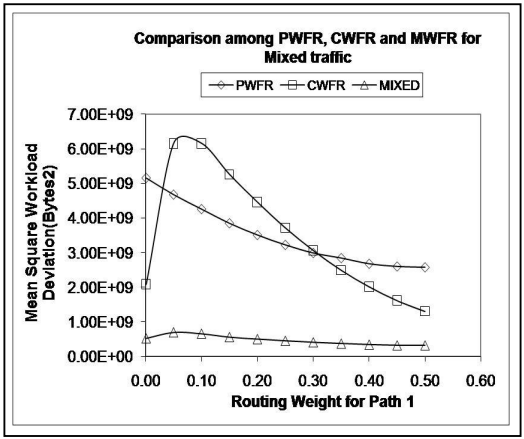
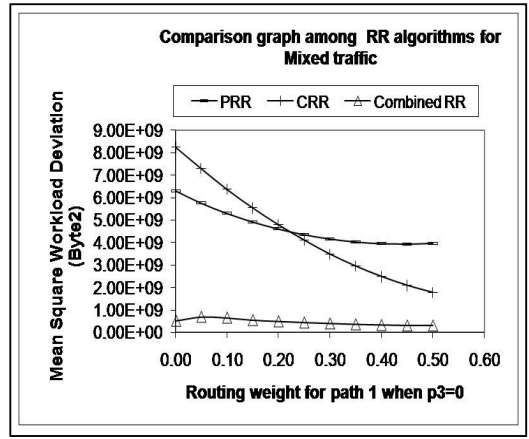

Figure 13. Comparison graph of PWFR, CWFR and MWFR for mixed traffic

Figure 14. Comparison graph for PGRR, CGRR and MRR

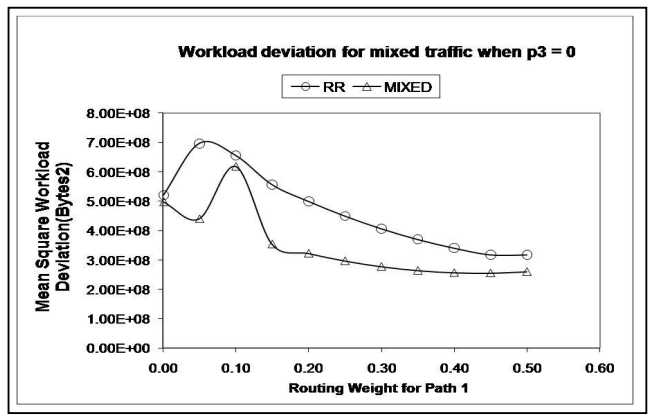

Figure 15. Comparison between MRR and MWFR for mixed traffic

### 3.5 Extended Simulation and Analysis

For examining the feasibility, robustness and stability of the different algorithms, we repeated the experiments but increased the number of outgoing links of a traffic splitter node and assigned random weight vectors to the multiple paths so that summation of total weight must be 1. Figure 16 depicts such this topology in our simulation.

After extensive tests using various weight vector combinations we found that MWFR works better than other algorithms. The result compares the effectiveness of different traffic splitter under three different traffic condition i.e. connection oriented, connectionless and mixed traffic whereas connection oriented traffic consist of traffic from transmission control protocol (TCP) connection only while connectionless traffic come from non-TCP connections. Mixed traffic is composed of both connection oriented and connectionless traffic.

From Figure 17 and 18 we conclude that PWFR works better than PGRR when the traffic is UDP and on the contrary, if traffic consists of TCP, CWFR works better than CGRR when the number of multipath of node 1 is 5. But when there is a mixture of both TCP and UDP traffic MWFR works better than MRR which is shown in Figure 19. So in a nutshell we can conclude that the extended simulation results demonstrates once again the superiority of MWFR algorithm over the others algorithms.

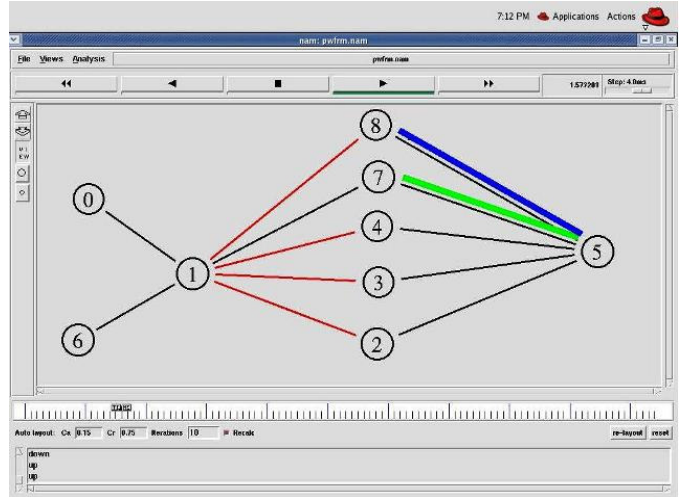

Figure 16. Simulated multipath network when number of path is 5

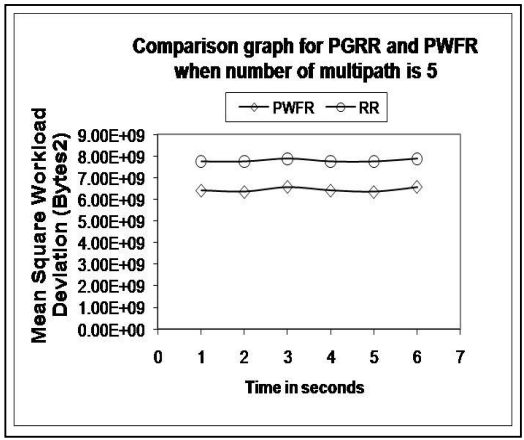 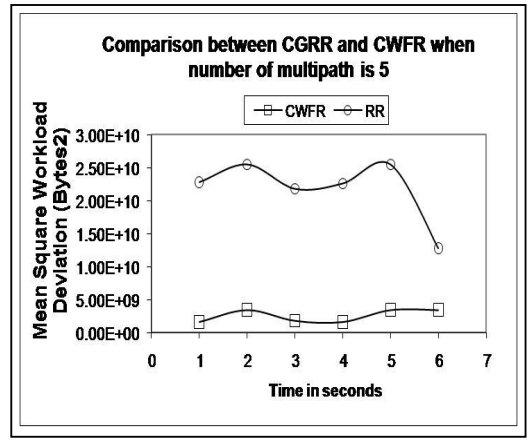

Figure 17. Comparison of PGRR and PWFR for UDP traffic when number of path is 5

Figure 18. Comparison of CGRR and CWFR for TCP traffic when number of path is 5

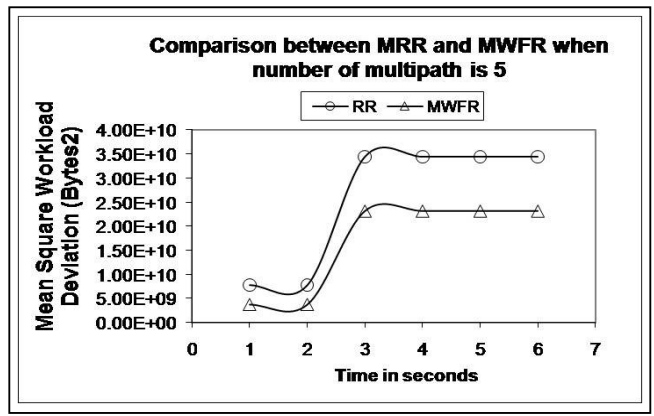

Figure 19. Comparison between MRR and MWFR for both UDP and TCP traffic when number of path is 5

## 6. CONCLUSION

The purpose of our work is to evaluate and enhance the performance of packet switching network for generalized load sharing and this is done by introducing some routing algorithms which reduces routing control traffic packets overhead and minimize workload deviation.

After extensive research and data analysis we conclude that as a traffic splitting algorithm divide traffic according to the given routing weight vector, hence the algorithm which keeps the mean square workload deviation as small as possible is the best one. Moreover, as real network traffic is a combination of TCP and UDP traffic, MWFR is the best choice as a traffic splitter for a node of multipath network environment. Again, after extensive simulation of our research work with varying network topology and changing the number of multipath, we are able to get stable result, so it can be concluded that for the number of multipath three is a satisfactory number for creating a multipath network environment.

There are many ways in which the current research work can be enhanced or expanded. The work is a simulation based work and not tested in real network. There is a future scope of research to see how it works for the real network. Also as this simulation work is very computation intensive; it is worthwhile to explore methods to improve simulation efficiency so that more accurate simulated results can be acquired.